\documentclass[
 reprint,
  twocolumn,
  prl,
  aps,
  superscriptaddress,
  showpacs,preprintnumbers,
  amsmath,amssymb,
floatfix,
]{revtex4}

\usepackage{graphicx,color}
\usepackage{multirow}

\begin{document}

\preprint{}

% \title{Spontaneous symmetry breaking in suspended bilayer graphene}
% \title{Gapped ground state in suspended bilayer graphene without an external applied field}
% \title{Spontaneously formed gapped ground state in suspended bilayer graphene}

\title{Spontaneously gapped ground state in suspended bilayer graphene}

\author{F. Freitag}
\affiliation{Department of Physics, University of Basel,
Klingelbergstrasse 82, CH-4056 Basel, Switzerland}

\author{J. Trbovic}
\affiliation{Department of Physics, University of Basel,
Klingelbergstrasse 82, CH-4056 Basel, Switzerland}

\author{M. Weiss}
\affiliation{Department of Physics, University of Basel,
Klingelbergstrasse 82, CH-4056 Basel, Switzerland}

\author{C. Sch{\"o}nenberger}
\email{Christian.Schoenenberger@unibas.ch} \affiliation{Department of
Physics, University of Basel, Klingelbergstrasse 82, CH-4056 Basel,
Switzerland}

\date{\today}

\begin{abstract}
Bilayer graphene bears an eight-fold degeneracy due to spin, valley and layer symmetry,
allowing for a wealth of broken symmetry states induced by magnetic or electric fields, by strain,
or even spontaneously by interaction. We study the electrical transport in clean current annealed suspended
bilayer graphene. We find two kind of devices. In bilayers of type B1 the eight-fold zero-energy Landau level (LL) is partially lifted above a
threshold field revealing an insulating $\nu=0$ quantum Hall state at the charge neutrality point (CNP).
In bilayers of type B2 the LL lifting is full and a gap appears in the differential conductance even
at zero magnetic field, suggesting an insulating spontaneously broken symmetry state.
Unlike B1, the minimum conductance in B2 is not exponentially suppressed, but remains finite
with a value $G \alt e^2/h$ even in a large magnetic field. We suggest that this phase of B2 is insulating
in the bulk and bound by compressible edge states.
\end{abstract}

% 72.80.Vp Electronic transport in graphene
% 73.43.Qt Magnetoresistance under: 73.43.-f Quantum Hall effects
% 73.23.-b Electronic transport in mesoscopic systems
% 73.22.Pr Electronic structure of graphene
% 73.22.Gk Broken symmetry phases

\maketitle

% =============================================================================================================
%\subsection*{Introduction}
% =============================================================================================================

Two dimensional electron systems (2DES) can host a large variety of ground states. Celebrated examples are the fractional quantum-Hall
effect \cite{Tsui1982,Du2009,Bolotin2009} and the Wigner crystal \cite{Wigner1934}, both being driven by Coulomb interaction. Bilayer
graphene
% with its strongly coupled layers of Dirac electrons
provides a further class of interacting 2DES~\cite{DasSarma2010}. In contrast to single layer graphene, the chiral charge
carriers are massive due to the coupling between the two layers~\cite{McCann2006a,Novoselov2006}. Owing to the large number of
symmetries, a wealth of ground states has been
predicted~\cite{Min2008,Nandkishore2010,Zhang2010,Nandkishore2010b,Gorbar2010,Zhang2011}.
%
% The possibility of opening a gap in the low-energy band structure by the application of a perpendicular
% electric field \cite{McCann2006b,Oostinga2008,Weitz2010} makes bilayer graphene a viable candidate for transistor technology.
% By applying such a field, resistances of 20~M$\Omega$ per square were reached in suspended bilayer graphene devices~\cite{Weitz2010}.
% The appearance of a quantum Hall state with diverging resistance at filling factor $\nu=0$ is a clear demonstration of the lifting of
% the eightfold degenerate zero-energy Landau level~\cite{Feldman2009,Weitz2010,Zhao2010}. So far, an external
% magnetic field was necessary to approach this insulating state.
%

Bilayer graphene proves to be interesting in terms of electron-electron interaction. In comparison with single layer
graphene the interaction parameter $r_{s}$ is about $30$ times higher and proportional to $1/\sqrt{n}$, where $n$ is the
charge carrier density~\cite{DasSarma2010}. The Coulomb interaction can even further be increased by suspending the sample.
% Due to the different environment of graphene, now vacuum on both sides instead of SiO$_2$ on one side, the dielectric polarizability is reduced.
Furthermore, cleaner devices can be obtained by current annealing~\cite{Moser2007,Bolotin2008}.
The subsequently lower disorder potential enables one to reach a lower
minimal carrier concentration $n$ at the charge neutrality point (CNP). At and around the CNP, electron-electron interaction
has been predicted to be able to spontaneously open a gap~\cite{Min2008,Nandkishore2010,Zhang2010,Nandkishore2010b,Gorbar2010,Zhang2011}.
This is opposed to an induced gap from the application of an external field \cite{McCann2006b,Oostinga2008,Weitz2010}
or mechanical strain~\cite{Choi2010,Falko2011}.

% In a quantizing magnetic field, the Landau levels (LL) lie at energies $E_{N} = \pm \hbar \omega_{c} \sqrt{N(N-1)}$ in
% bilayer  graphene, where $\omega_{c}=eB/m^{*}$ is the cyclotron frequency and $N=0,1,2,\ldots$ and hence there are two
% orbitals, $E_{0}$ and $E_{1}$,  at zero energy \cite{McCann2006a,Novoselov2006}.
Bilayer graphene has an eightfold degenerate LL at zero energy.
As the Hall conductivity is quantized at values of $\sigma_{xy} = \nu \cdot e^{2}/h$, where the filling factor
$\nu$ is given by $\nu = \pm 4(N+1)$, a step of 8$e^{2}/h$ is observed from
$\nu=-4$ to $\nu=4$ around the CNP~\cite{Novoselov2006}. The eightfold zero-energy LL degeneracy can be lifted. For example,
the Zeeman energy is able to break the spin symmetry.
This manifests itself in a partial lifting with a quantum Hall plateau appearing at $\nu=0$~\cite{Feldman2009}.
A breaking of symmetries can also be induced by strong electron-electron interaction \cite{Nomura2006,Gorbar2010}.
If all symmetries are lifted, quantum Hall plateaus appear at filling factors $\nu=0$, $\pm 1$, $\pm 2$, $\pm 3$, $\dots$. Magnetic fields of
\mbox{$30-45$\,T} were required to see this lifting in silicon dioxide supported devices, for both single~\cite{Zhang2006} and bilayer
graphene~\cite{Zhao2010}, until Feldman \textit{et al.} succeeded in observing the effect at low magnetic fields in
suspended bilayer graphene \cite{Feldman2009}.

The most striking state is the $\nu=0$ state, whose nature is under debate
for both single layer~\cite{Abanin2007,Checkelsky2009,Zhang2010a} and bilayer graphene~\cite{Feldman2009,Weitz2010,Zhao2010}.
For bilayer graphene, several possibilities are being discussed, such as the quantum Hall ferro\-magnet (QHF)~\cite{Nomura2006,Gorbar2010},
the quantum anomalous Hall insulator (AHI)~\cite{Raghu2008,Nandkishore2010b}, or a ferroelectric phase \cite{Zhang2010}.
Using differential conductance spectroscopy, we find a new class of bilayer samples, which are evidentially gapped at
the CNP in zero magnetic and electric field.

% =============================================================================================================
%\subsection*{Experimental, Samples}
% =============================================================================================================

Suspended graphene devices were fabricated by mechanical exfoliation of natural graphite transferred to
a doped Si wafer with a $300$\,nm top SiO$_2$ layer. The number of graphene layers was determined by Raman spectroscopy.
The devices were then annealed for several hours in vacuum ($10^{-7}$\,mbar) at $200$\,$^{\circ}$C before the
electrical contacts made from Cr/Au ($1$/$70$\,nm) bilayers were fabricated by electron-beam lithography.
Thereafter, SiO$_{2}$ was etched in buffered hydrofluoric acid (HF). After mounting a device into a $^{3}$He
cryostat, we performed current annealing by applying a DC current at $1.5$\,K.
This procedure was repeated with higher currents until the electrical conductance $G(V_g)$ as a function of the
gate voltage $V_g$ applied to the doped substrate showed a pronounced dependence with a charge neutrality point (CNP), where
$G$ has a minimum, close to $V_g=0$, reminiscent of a high quality device (this usually required current densities of
up to $8\cdot 10^{7}$\,A/cm$^{2}$). Conductance measurements were carried out with a lock-in amplifier applying a $20$\,$\mu$V AC
voltage onto which a DC bias voltage could be superimposed.

\begin{figure}[ht!]
\begin{center}
  \includegraphics[width=1.0\linewidth]{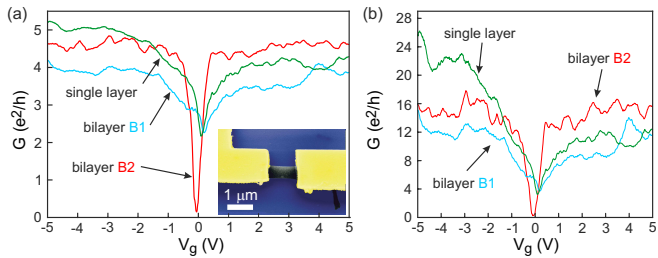}
\end{center}
  \caption{(Color online)
  % \textbf{Gate response of suspended graphene devices.}
  (a) Comparison of the measured dependence of the conductance $G$
  on the applied gate voltage $V_{g}$ for three suspended devices:
  monolayer graphene (length$\times$width 1$\times$2~$\mu m^2$, $T=2$~K),
  bilayer type B1 (2$\times$0.8~$\mu m^2$, $T=230$~mK), and bilayer type B2
  (1$\times$1.5~$\mu m^2$, $T=230$~mK). In (b) the contact resistance has been subtracted.
  } \label{GvsBG}
\end{figure}

Fig.~1 shows representative measurements of the two-terminal conductance $G$ of  suspended graphene devices
% (see inset of Fig.~1)
when $n$ is altered by the back-gate voltage $V_{g}$. The CNP is close to $V_{g}=$~0~V for all samples,
indicating that only few charged impurities reside on the graphene.
% The typical minimum conductivity $\sigma_{min}$ of single layer and bilayer graphene is expected to be of the order of $e^{2}/h$
% \cite{DasSarma2010,Trushin2010}. This is well fulfilled by the single layer sample with \mbox{$G_{min}=$ 2.1 $e^{2}/h$},
% where $G_{min}=(W/L) \cdot \sigma_{min}$, and the bilayer device of type B1, where \mbox{$G_{min}=$ 2.3 $e^{2}/h$} (sample
% dimensions listed in Fig.~\ref{GvsBG}).
Both single layer and bilayers of type B1 display a smooth transition from low $G$ at the CNP to higher $G$ values at larger
$n$, as expected from the V-shaped conductances found in recent literature~\cite{Morozov2008,Du2008,Feldman2009}.
In contrast, bilayer samples of type B2 are very low conducting at the CNP with \mbox{$G_{min} <$ 0.2 $e^{2}/h$} at $230$\,mK,
which is considerably lower than in previous reports~\cite{Weitz2010}. Furthermore, as the gate voltage is tuned away from the
CNP, $G$ increases sharply and then quickly saturates for $|V_{g}| > 0.5$\,V. Note, that this is even the case, when the
contact resistance is subtracted as shown in Fig.~1b.

% =============================================================================================================
%\subsection*{Hall effect measurements}
% =============================================================================================================

\begin{figure}[ht!]
\begin{center}
  \includegraphics[width=1.0\linewidth]{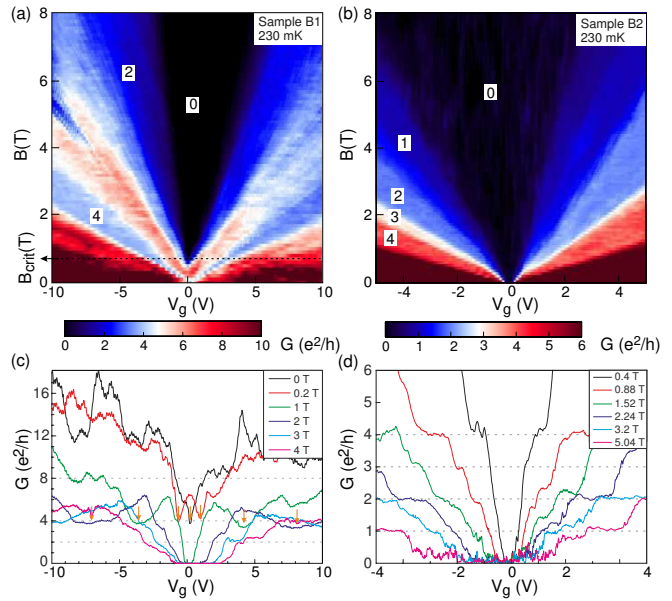}
\end{center}
  \caption{(Color online)
  % \textbf{Two-terminal quantum Hall effect.}
  The dependence of the linear conductance $G$ on the gate voltage $V_g$ and perpendicular magnetic field $B$ reveal two type
  of samples. In sample B1, shown in (a,c) the plateaus at filling factors $\nu=0$, $\pm 2$ and $\pm 4$ are well developed.
  The $\nu=\pm 4$ plateau (minima in (c), see arrows) extends to very low magnetic fields. For sample B2, a full
  lifting of the eightfold Landau level degeneracy is observed, as plateaus at odd fillings appear as well. The curves
  in (c,d) show lines cuts at constant $B$. Appropriate contact resistances were subtracted.
  } \label{QHE}
\end{figure}

When placed in a perpendicular magnetic field $B$, samples B1 and B2 reveal substantially different quantum Hall features,
as shown in Fig.~2. As the measurements were performed in a two-terminal configuration, they include a
contact resistance~\cite{Williams2009}. We determine the contact resistance by
matching $G$ to the closest integer value of the quantized Hall conductance (supplemental material~\cite{supplemental})
and then subtracted it from the data. First, we discuss sample B1 and then sample B2.

In sample B1 we observe a partial lifting of the eightfold zero energy LL degeneracy, leading to the $\nu= \pm 2$ and $0$
states above a critical magnetic field of $B_{crit} \approx 0.75$\,T (Fig.~2a). In the magnetic field range
$0\leq B \leq 0.75$\,T we observe the same Hall sequence as in conventional devices where the conductance has a
step of $8 e^{2}/h$ from $\nu=-4$ to $+4$. When applying $B > B_{crit}$ an insulating state emerges
around the CNP, followed by the $\nu=\pm 2$ state with a two-fold degeneracy remaining. We also note that the $\nu=\pm 4$
state appears to extend all the way down to the CNP at zero magnetic field~\cite{Martin2010,Nandkishore2010b}. The
corresponding line cuts from the color scale are shown in Fig.~2c to illustrate the evolution of the CNP into the
$\nu=\pm 4$ state at low fields and the appearance of the broken symmetry states $\nu=0$ and $\pm 2$ at higher fields~\cite{Feldman2009}.
Unlike sample B1, B2 shows a fully lifted zero-energy LL, manifesting in the appearance of Hall plateaus for
odd filling factors $\nu$. In analogy to sample B1, we label the low conducting region around the CNP in sample B2 with \mbox{$\nu=0$},
although this state maintains a finite conductance as we will discuss below.

\begin{figure}[ht!]
\begin{center}
  \includegraphics[width=1.0\linewidth]{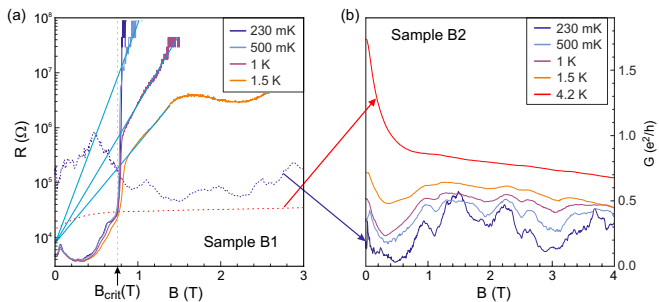}
\end{center}
  \caption{(Color online)
  % \textbf{Phase transition and minimum conductance in magnetic field.}
  Difference in resistance $R$ (a) of sample B1 and conductance $G$ (b) of sample B2 as a function of magnetic field $B$ at the CNP.
  A sharp transition to an insulating $\nu=0$ state appears at $B_{crit}= 0.75$\,T (arrow) in B1. In this state, $R$ is thermally
  activated~\cite{supplemental} with an activation energy $\Delta E$ proportional to $B$. The lines are guides to the eyes for this dependence.
  For comparison we show the resistance data for sample type B2 (dotted) within the
  same graph. In contrast to sample B1, sample B2 does not show a field-induced transition to an insulating state.
  At low temperatures, $G$ is remarkably insensitive to $B$, but displays a relatively low conductance value $ < e^2/h$.
  Appropriate contact resistances are subtracted.
  } \label{TransitionInB}
\end{figure}

In the following Fig.~3 we investigate the properties of the low conducting state at $\nu=0$ at low charge
carrier density as a function of $B$ and $T$ for both samples. For device B1 (Fig.~3a), we find that at low $B$
the resistance $R$ at the CNP remains around \mbox{$R=6$\,k$\Omega$}, but when a critical perpendicular magnetic field
of $B_{crit}\approx 0.75$\,T is reached, it increases sharply to $10^{8}$\,$\Omega$, the maximum resistance that our measurement
set-up can resolve. This behavior can be attributed to the formation of a quantum Hall state at $\nu=0$~\cite{Checkelsky2009}.
%
% The fact that this state is insulating in electrical measurements can be understood by looking at the conductivities,
% which on a Hall plateau are given by $\sigma_{xy}=\nu \cdot e^{2}/h$ and $\sigma_{xx} \rightarrow 0$. Taking the conversion to
% resistivity $\rho_{xx}=\sigma_{xx}/(\sigma_{xx}^2+\sigma_{xy}^2)$ and setting $\nu=0$, $\rho_{xx}\rightarrow\infty$ follows,
% which is what we observe for $B > B_{crit}$.
%
Since the Fermi energy now lies in between two Landau levels, a thermally activated behavior is expected for
the electrical resistance $R$. This is confirmed in the experiment,
revealing a strong dependence of $R$ on temperature $T$ above the critical field following the law
$R \propto exp(\Delta E/2k_BT)$~\cite{supplemental}. The activation energy $\Delta E$ is linearly dependent on $B$ with a
value that amounts to $1.1$\,meV/T $=13$\,K/T.
% In a non-interacting picture of spin\-less electrons $\Delta E$ would be determined by the Landau level separation
% $\hbar\omega_c$. Using this ansatz, yields a an effective mass of $m^{*}\approx 0.1$\,$m_e$. Though this is a reasonable mass
% for bilayer graphene, the non-interacting picture is not appropriate, since it predicts the eightfold degenerate zero-energy LL.
Note, that the spin Zeeman splitting is much smaller, only amounting to $\sim 0.7$\,K/T.
% The observed large activation energy has to be explained by interaction which is causing the symmetry breaking.
Feldman~{\it et al.}~\cite{Feldman2009} deduce in their experiment $\Delta E = 3.5 - 10.5$\,K/T, which is
somewhat lower than our number. In a recent theory, taking interactions into account, the energy gap of the $\nu=0$ state has been
calculated to be $14.3$\,K/T~\cite{Gorbar2010}.
% This number is in very good agreement with our experiment.

The dotted curves in Fig.~3a show the resistance of B2 in direct comparison to that of B1 at $230$\,mK and $4$\,K.
We find that at $B=0$\,T B2 has an order of magnitude higher resistance than B1, whereas at higher magnetic fields, B1
is several orders more resistive. Fig.~3b elaborates on the conductance of B2 at the CNP as a function of
$B$ and temperature. Most notably, $G$ at $B \agt 1$\,T is nearly independent of $B$ with the exception of
fluctuations most likely due to localized states~\cite{Bolotin2009}.

The marked differences in the magnetic field dependence clearly demonstrate that sample B1 and B2 differ. In sample B1, the
LL degeneracy is partially lifted for $B > B_{crit}$ $\approx 0.75$\,T. This lifting includes a transition into a
$\nu=0$ quantum Hall plateau in the vicinity of the CNP. On the other hand, sample B2 reveals a fully lifted LL, where all
Hall plateaus appear already a small $B \sim 1$\,T. Furthermore, sample B2 stays conductive at the CNP even at higher
$B$ of up to $8$\,T. We note, that sample B1 is similar in characteristics to the one reported by Feldman~{\it et al.}~\cite{Feldman2009},
whereas B2 shows new features.

\begin{figure}[ht!]
\begin{center}
  \includegraphics[width=1.0\linewidth]{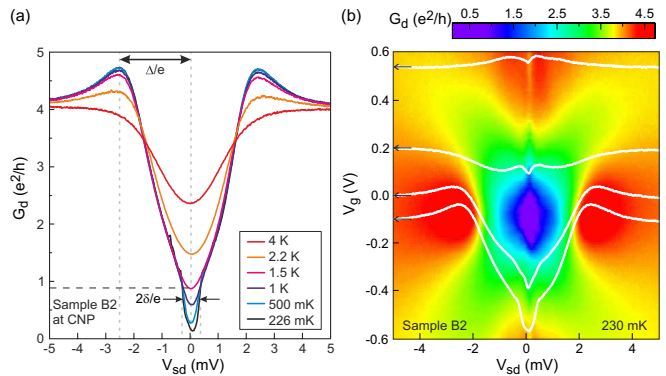}
\end{center}
  \caption{(Color online)
  % \textbf{Gaps in bilayer graphene}
  (a) Temperature dependence of the differential conductance $G_d$ around the
  CNP of sample B2 as a function of the source-drain bias voltage $V_{sd}$. Two gaps with size \mbox{$\Delta=$ 2.5 meV} and
  \mbox{$\delta=$~0.35 meV} appear that both display a distinct temperature dependence.
  % The smaller gap $\delta$ becomes relevant at \mbox{$T<$ 1 K} and reduces $G$ below 0.7~$e^{2}/h$.
  % Consequently, the larger gap $\Delta$ saturates between $1.5$\,K and $1$\,K at \mbox{$G \approx$ 0.7~$e^{2}/h$}.
  % This gap remains visible even at 4~K. Further,  the sizes of the gaps \mbox{$\Delta=$ 2.5 meV} and
  % are indicated by the dotted lines.
  (b) Color-scale of $G_d$ as a function of $V_{sd}$ and gate voltage $V_{g}$ at $230$\,mK.
  Line cuts are taken at the values of $V_{g}$ marked by the arrows.} \label{Gap}
\end{figure}

We further investigate the nature of sample B2 by measuring the differential conductance $G_d$ as a function of the applied DC
bias $V_{sd}$ between source and drain contacts at $B=0$\,T at the CNP ($V_{g}=-0.1$\,V), where $G_d$ is suppressed.
Fig.~4a summarizes the findings for different temperatures from $226$\,mK to $4$\,K (no contact resistance subtracted).
Two gaps can clearly be identified:
When going from large source-drain bias $V_{sd} > 4$\,mV towards small voltages, the larger gap $\Delta$ sets in at
$V_{sd}=\pm 2.5$\,mV, where $G_d$ is decreased from around $4 e^{2}/h$ to $0.9 e^{2}/h$ in the data measured at $226$\,mK.
The smaller gap $\delta$ appears at voltages $|V_{sd}|\alt 0.35$\,mV and it reduces $G_d$ from $0.9 e^{2}/h$ to less than
$0.2 e^{2}/h$. By increasing the temperature from $226$\,mK on, the smaller gap $\delta$ is first reduced and then vanishes at $1$\,K.
% Meanwhile, the larger gap $\Delta$ saturates at temperatures below 1~K and remains open even at 4~K, although $G_{min}$ is now
% significantly higher. For a source-drain voltage of more than 4~mV the conductance starts to converge to a value independent
% of the temperature in the range of our measurement.

In order to identify the origin of these two gaps, a color scale plot of the differential conductance $G_d$ against $V_{sd}$
and the gate voltage $V_g$ at $230$\,mK is shown in Fig.~4b. The line cut at the CNP ($V_{g}=-0.1$\,V) shows again the
two gaps in electron transport. As $V_g$ and thus the charge carrier concentration is increased, the two gaps exhibit distinct
changes. The larger gap $\Delta$ disappears, while the smaller gap $\delta$ still exists in the metallic graphene
regime at $V_{g} >0.5$\,V but the relative dip is less pronounced.
This behavior is qualitatively consistent with Coulomb charging of the whole flake~\cite{Devoret1990}.
We estimate a single-electron charging energy of $1$\,meV for a flake of width $1.5$\,$\mu$m.
Because the contact conductances of $\sim 4-8$\,$e^2/h$ are substantially larger than $e^2/h$, charge
quantization is only weak and no strong Coulomb blockade gap is expected. One rather expects the conductance to
display a `weak' conductance suppression by something like $25-50$\,$\%$ around zero bias, in agreement with the observation.
In contrast to the small gap, the larger gap $\Delta$ is strongly dependent on the charge carrier density. At the CNP, it has its maximum
magnitude, but only slightly away it starts to close. The line cut at $V_{g}=0.2$\,V already bears little sign of the gap
$\Delta$. We therefore conclude that it must be a feature intrinsic to the low-energy band structure of bilayer graphene and
that this gap is formed spontaneously at zero magnetic and at \emph{zero} electric field.
We emphasize that the electric field induced by the back-gate voltage in the vicinity of the gap feature
of $V_g\approx 100$\,mV is negligible. In the tight-binding band structure calculation of McCann~\cite{McCann2006b}
an induce a gap of $\Delta=2.5$\,mV, as we observe it in our experiment, would require a back-gate voltage of at least $10$\,V.
%
% To summarize, a partial lifting of the LL degeneracy is observed in sample B1 above a critical magnetic field
% $B_{crit}\approx 0.75~T$. This lifting includes the appearance of a $\nu=0$ quantum Hall state~\cite{Feldman2009,Zhao2010}.
% In contrast, sample B2 shows a complete lifting of the zero-energy LL with no apparent onset, suggesting a symmetry breaking
% that appears spontaneously already in zero field. This picture is supported by the observed gap feature at zero magnetic field.
%
The $\Delta$ gap can be associated with the $\nu=0$ state, as the low-conductance region in the color scale plot of
Fig.~\ref{QHE}b extends from large magnetic fields all the way down to zero magnetic field with no apparent phase
boundary~\cite{supplemental}, distinctly different to the finding of Weitz~{\it et al.}~\cite{Weitz2010}.
Although sample B1 and B2 have a $\nu=0$ state around the CNP in our argument, these states
are electrically different. In sample B1, the resistance evidently increases to infinity, whereas it stays finite in B2.
This difference can be explained by insulating phases, differing in their edge-state structure~\cite{Raghu2008,Nandkishore2010b,Zhang2011}.

Sample B1 has two phases, a low-magnetic field phase and a broken symmetry state induced by a small magnetic
field of $B > B_{crit}$. The latter most likely is a quantum Hall ferro\-magnet~\cite{Nomura2006,Gorbar2010,Nandkishore2010b}.
The phase at low magnetic field has been assigned to a gapped anomalous Hall insulator (AHI)
in which topologically protected edge states should provide a conductance of up to
$4e^2/h$~\cite{Martin2010,Nandkishore2010b,Zhang2011}. This scenario is somewhat supported by the quantum Hall states at
$\nu\pm 4$ that persist all the way down to $B=0$ (see arrows in Fig.~2c). A similar observation has been made in
compressibility measurements by Martin et al.~\cite{Martin2010}. Because B2 is the cleaner sample of the two~\cite{comment},
we rather think that low-field phase of B1 is a normal state, not a broken symmetry state.
In contrast, the low-density phase of B2 is a broken symmetry state with edge states.
If we subtract the small gap $\delta$ in sample B2, the measured conductance $G$ is $\approx 0.8$\,$e^2/h$,
which is smaller than the ballistic channel conductance of any gapped phase with edge states.
This suggests that the gapped phase is either not single domain
or that the edge states are not topologically protected, allowing for partial back-scattering.
Further work is needed to determine the nature of the edge states and assign it to
broken electron-hole, valley or spin-symmetry~\cite{Abanin2006,Martin2008,Nandkishore2010b,Zhang2011}.

In conclusion, using differential conductance spectroscopy we found a new type of bilayer whose spectral density is
gapped at zero magnetic and zero electric field. Though this state is due to an insulating phase, the non-vanishing
conductance $\approx 0.8$\,$e^2/h$, which is surprisingly robust in magnetic field, suggests that edge states are present.

This work was financed by the Swiss NSF, the ESF programme Euro\-graphene, the NCCR Nano and the NCCR Quantum. We
acknowledge access to Raman microscope provided by C. Stampfer and C. Hierold. We are also grateful to A. Baumgartner, D.
Maslov, A. Morpurgo, A. Yacoby, V. Fal'ko and L. Levitov for discussions.


\begin{thebibliography}{10}
\bibitem{comment}
  We can convert B2 into B1 by exposing the sample to air. To
  convert B1 into B2 requires current annealing at low temperatures.
\end{thebibliography}
\end{document}